\documentclass[%
 reprint,
superscriptaddress,
longbibliography,
 amsmath,amssymb,
 aps,
 prb,
]{revtex4-1}

\usepackage{sidecap}
\sidecaptionvpos{figure}{c}
 \usepackage[T1]{fontenc}

\usepackage{graphicx}
\usepackage{dcolumn}
\usepackage{bm}
\usepackage[
colorlinks=true,
linkcolor=blue,
citecolor=blue,
filecolor=blue,
urlcolor=blue]{hyperref}

\usepackage{amsmath,amsfonts,
	amsgen,amsbsy,amsopn,amstext,amssymb
}
\usepackage{epstopdf}
\usepackage{xcolor}

\usepackage{soul} 

\renewcommand{\i}{i}

\definecolor{redish}{RGB}{255, 173, 231}
\definecolor{bluish}{RGB}{205, 224, 247}

\begin{document}

\preprint{}

\title{Stealth acoustic materials}
\author{V. Romero-Garc\'ia}\email{vicente.romero@univ-lemans.fr}
\author{N. Lamothe}
\author{G. Theocharis}
\author{O. Richoux}
\affiliation{Laboratoire d'Acoustique de l'Universit\'e du Mans, LAUM - UMR 6613 CNRS, Le Mans Universit\'e, Avenue Olivier Messiaen, 72085 LE MANS CEDEX 9, France}

\author{L. M. Garc\'ia-Raffi}
\affiliation{Universitat Polit\`ecnica de Val\`encia, Cam\'i­ de vera s/n, 46022, Val\`encia, Spain}

\date{\today}

\begin{abstract}
We report the experimental design of a 1D stealth acoustic material, namely a material that suppresses the acoustic scattering for a given set of incident wave vectors. The material consists of multiple scatterers, rigid diaphragms, located in an air-filled acoustic waveguide. The position of the scatterers has been chosen such that in the Born approximation a suppression of the scattering for a broad range of frequencies is achieved and thus a broadband transparency. Experimental results are found in excellent agreement with the theory despite the presence of losses and the finite size of the material, features that are not captured in the theory. This robustness as well as the  generality of the results motivates realistic potential applications for the design of transparent materials in acoustics and other fields of wave physics.
\end{abstract}

\pacs{Valid PACS appear here}
\keywords{Stealth materials, Transparency, Hyperuniform}
\maketitle

\section{Introduction}\label{sec:intro}
Wave scattering represents one of the most studied phenomena in wave physics. When a system with refractive index contrast between the obstacle and the surrounding material is radiated by an incident wave, the incident energy is both radiated and absorbed by the obstacle, forming a scattering pattern which strongly depends on the geometry and size of the obstacle as well as on the frequency-dependent properties of its constituents. The control of the scattering of waves represents a major topic of interest in acoustics\cite{Shen14, Jimenez17b}, photonics\cite{Shen15} and electromagnetism\cite{Asadchy15} due to the vast possibilities in fundamental physics with potential applications. Complex media\cite{martin06} and recently metamaterials\cite{Engheta06}, have been shown as potential candidates to control the scattering of different types of waves. In particular it has been shown that metamaterials in both electromagnetism\cite{Alu05} and acoustics\cite{Chen11}  can be used to induce a dramatic drop in the scattering cross section of cylindrical and spherical objects, making them nearly invisible or transparent to an outside observer.

Ordered media, known as photonic\cite{Yablonovitch87, John87, joannopoulos08} or phononic crystals\cite{Sigalas92, Martinez95, Deymier13} in optics and acoustics respectively, have been exploited due to their particular dispersion relation, presenting band gaps for different ranges of frequencies. In periodic structures, order is present on every scale, allowing for multiple Bragg scattering and localized states to form a band gap when the wavelength of the propagating wave is of the order of a multiple of the distance between scatterers. In the Born approximation, periodic structures are transparent at low frequencies\cite{kittel04}. Among other potential applications, these systems have motivated tunable frequency filters\cite{Sanchez98, Romero11a}, beam forming devices\cite{Perez07}, waveguides\cite{Khelif04}, wave traps\cite{Sigalas98, Khelif03, Romero10a} and slow wave systems\cite{Baba08, Kaya12, Theocharis14, Groby16}. 

On the other hand, disordered media are also good alternatives to control wave scattering in a different manner\cite{ishimaru97}. Waves entering in a disordered material are scattered several times before exiting in random directions, producing interferences leading to interesting, and sometimes unexpected, physical phenomena\cite{Wiersma}. For example, in the acoustical\cite{Hu08} or optical\cite{Sperling13} analogy of Anderson localization, multiple scattering creates modes with a high level of spatial confinement in which the wave remains in the sample for a long time. Contrary to the periodic case, configurations of multiple scatterers with random and independent coordinates have no order at any scale and as a consequence no characteristic length appears in the system. Therefore no band gap is formed in such kind of structured materials and wave is randomly scattered.

Other approaches based on complex structures with correlated disorder\cite{Fan91, Kuhl00, Torquato02, Uche04, Kuhl08, Batten08, Florescu09, Dietz12, Man13a, Man13b, Torquato16, Torquato15, Shen15, Leseur16, Gkantzounis17}, created by introducing local correlations between positions of scatterers in an otherwise random ensemble, have been used to control the overall scattering strength of the material and create a strong wavelength dependence of the transmission. If the structure is properly designed, the system can even present isotropic band gaps\cite{Man13a, Man13b} in the absence of periodicity.  Such disordered materials are known as hyperuniform materials\cite{Torquato02, Torquato15} and nearly suppress all scattering at low frequencies. Hyperuniform materials are a subset of the systems known as stealth materials, referring to configurations of multiple scatterers that completely suppress scattering of incident radiation for a set of wave vectors, and thus, are transparent at these wavelengths. The hyperuniform materials represent the low-$k$ limit of the stealth materials and are, counterintuitively, disordered and highly degenerate. 

In this work we construct disordered stealth configurations by engineering the material properties in a controlled manner in such a way that the system prevents scattering only at prescribed wavelengths with no restrictions on any other wavelengths. In particular we focus on the suppression of the Bragg scattering of the corresponding periodic case and make the structured system transparent for a broadband range of frequencies. The inverse problem to generate the configuration of multiple scatterers with a specific scattering properties is non-trivial, as multiple solutions are conceivable. The approach is based on the optimization of the structure factor which, in the Born approximation, can be used as the outcome of a scattering experiment. The configuration of multiple scatterers is used to construct one dimensional (1D) stealth structures made of rigid diaphragms embedded in an air-filled waveguide. The system is analyzed by using the Transfer Matrix Method (TMM). Full wave numerical simulations have been done by using Finite Element Method (FEM). The system is analyzed considering the intrinsic viscothermal losses of the system. Experiments are performed to validate the theoretical results.

\section{Scattering by multiple scatterers}\label{sec:theory}

\subsection{Structure factor}
\begin{figure}
\centering
\includegraphics[width=85mm]{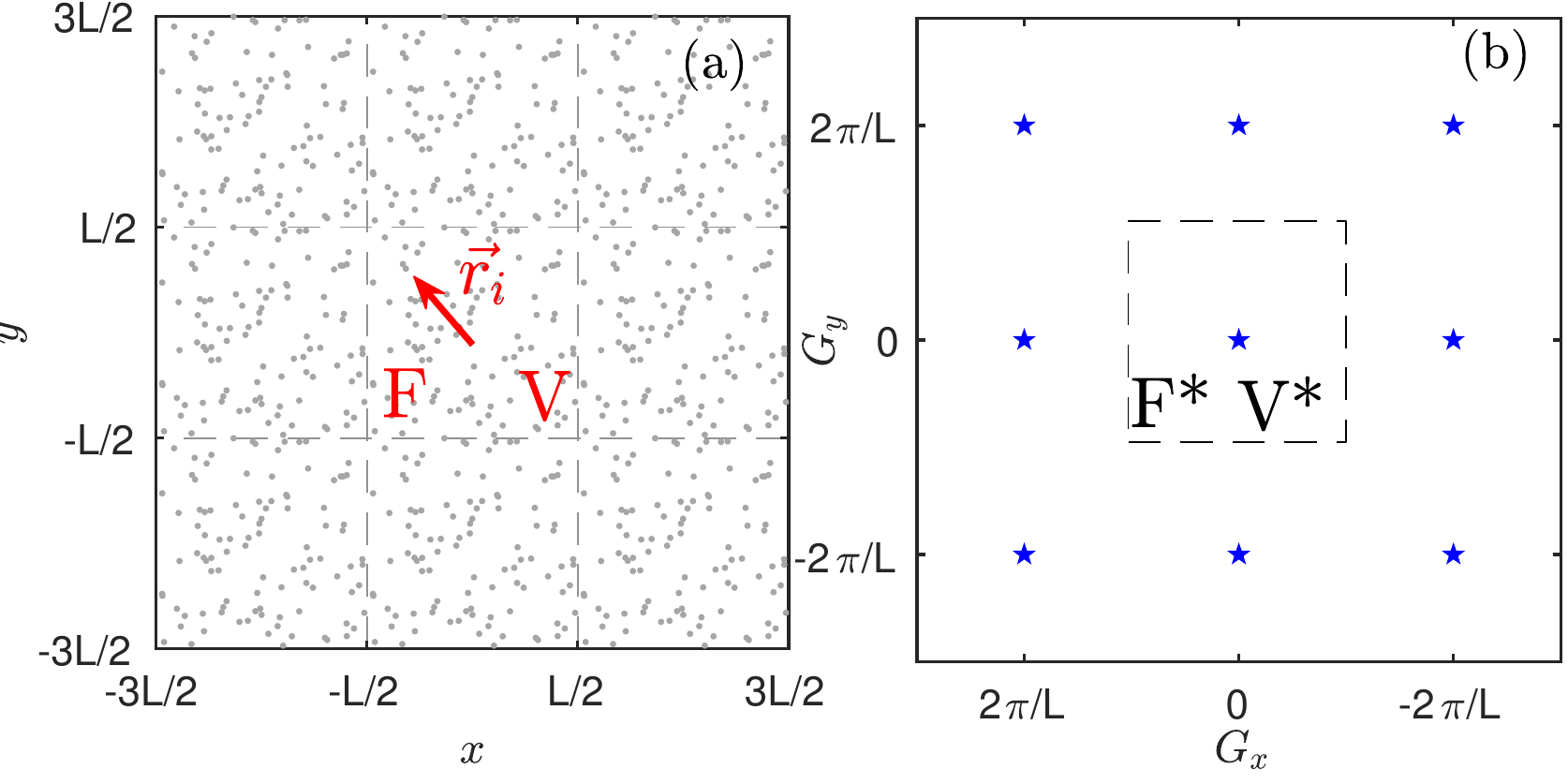}
\caption{\label{fig:fig2p} (Color online) Schematic example for the case of square periodicity in 2D with a lattice period $L$. (a) Direct space. (b) Reciprocal space.}
\end{figure}

We are interested in the study of the scattering of waves by disordered materials made of a discrete distribution of point scatterers. In order to theoretically and numerically deal with such distribution of scatterers, the space is periodically divided using a lattice $\bf{F}$ which unit cell has a volume $\bf{V}$ and in which $N$ scatterers are distributed at the positions $\vec{r}_i$ ($i=1,\ldots,N$) as schematically shown in a two dimensional example in Fig. \ref{fig:fig2p}(a). The lattice $\bf{F}$ has a reciprocal lattice $\bf{F^*}$ in which the lattice sites are specified by the reciprocal lattice vector, $\vec{G}$, with the property $\vec{G}\vec{R}=2\pi m$ for all lattice places $\vec{R}$, where $m=\pm1,\pm2,...$. The reciprocal fundamental unit cell has a volume $\mathbf{V^*}=(2\pi)^d/\mathbf{V}$ (where $d$ is the dimension of the $d$-dimensional Euclidean space in which the problem is defined). Although this choice introduces anisotropy, the periodic boundary conditions do not affect the generality of the properties discussed in this work. 

Let us consider the scattering of a beam with wavelength $\lambda$ by the assembly of $N$ scatterers. We assume that the scattering is weak, so that the amplitude of the incident beam is constant throughout the sample volume (Born approximation), and only scattered waves of the first order are considered; absorption, refraction and higher order scattering can be neglected (kinematic diffraction). The direction of any scattered wave is defined by its scattering vector $\vec{G}=\vec{k_{s}} -\vec{k_{0}}$, where $\vec{k_{s}}$ and $\vec{k_{0}}$ are the scattered and incident beam wavevectors, being $\theta$ incidence angle. For elastic scattering, $|\vec {k} _{s}|=|\vec {k_{0}} |=|\vec {k} |=2\pi /\lambda$ and then $G=|\vec {G} |={{\frac {4\pi }{\lambda }}\sin(\theta)}$. The amplitude and phase of this scattered wave will be the vector sum of the scattered waves from all the scatterers $\Psi _{s}(\vec {q} )=\sum _{i=1}^{N}f_{i}{e} ^{-\imath\vec {G} \vec {r} _{i}}$, with $f_i$ the atomic structure factor. The scattered intensity reads as
\begin{eqnarray}
I(\vec {G} )&=&\Psi _{s}(\vec {G} ).\Psi _{s}^{*}(\vec {G} )\nonumber\\
&=&\sum _{j=1}^{N}f_{j} {e} ^{-i\vec {G} \vec {r} _{j}}\times \sum _{k=1}^{N}f_{k} {e} ^{i\vec {G} \vec {r} _{k}}\nonumber\\
&=&\sum _{j=1}^{N}\sum _{k=1}^{N}f_{j}f_{k} {e} ^{-\imath \vec {G} (\vec {r} _{j}-\vec {r} _{k})}.
\end{eqnarray}
The structure factor, $S(\vec{G})$, is defined as this intensity normalized by $1/\sum _{j=1}^{N}f_{j}^{2}$
\begin{eqnarray}
S(\vec {G} )={\frac {1}{\sum _{j=1}^{N}f_{j}^{2}}}\sum _{j=1}^{N}\sum _{k=1}^{N}f_{j}f_{k}{e} ^{-\imath \vec {G} (\vec {r} _{j}-\vec {r} _{k})}.
\end{eqnarray}
If all the scatterers are identical, then
\begin{eqnarray}
I(\vec {G} )=f^{2}\sum _{j=1}^{N}\sum _{k=1}^{N}\ {e} ^{-\imath \vec {G} (\vec {r} _{j}-\vec {r} _{k})},
\end{eqnarray}
so
\begin{eqnarray}
S(\vec{G})={\frac {1}{N}}\sum _{j=1}^{N}\sum _{k=1}^{N} {e} ^{-\imath \vec{G} (\vec{r} _{j}-\vec{r} _{k})} = {\frac {1}{N}}\left|\sum _{j=1}^{N}{e} ^{\imath \vec{G} \vec{r} _{j}}\right|^2 .
\end{eqnarray}
Therefore the structure factor $S(\vec{G})$, is proportional to the intensity of scattering of incident radiation from a configuration of $N$ scatterers. It is worth noting here that the structure factor can be also related with the scattering cross section as follows
\begin{eqnarray}
\frac{d\sigma}{d\Omega}=f^{2}\sum _{j=1}^{N}\sum _{k=1}^{N}\ {e} ^{-\imath \vec {G} (\vec {r} _{j}-\vec {r} _{k})}=f^2NS(\vec{G}),
\end{eqnarray}
where $\sigma$ is the total cross section and $\Omega$ is the solid angle.

\begin{figure*}
\includegraphics[width=17.5cm]{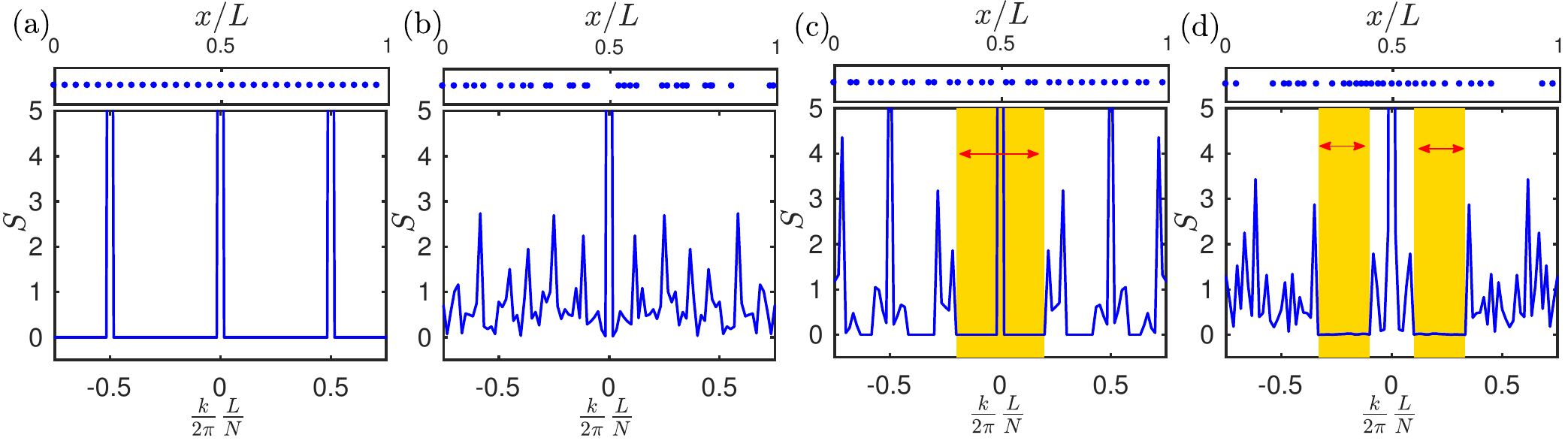}
\caption{(Color online) Representation of the structure factor for (a) a periodic, (b) a random, (c) a hyperuniform and (d) a stealth distributions of $N=30$ scatterers. The hyperuniform configuration is obtained to reduce scattering in the range $|k|\leq |k_B/2.5|$ (where $k_B=\pi N/L$ is the Bragg condition of the corresponding periodic configuration). The stealth configuration is optimized for $k_B/5\leq |k| \leq k_B/1.5$. The target range of frequencies of the two last cases are shown by the yellow area in (c) and (d).}
\label{fig:fig2}
\end{figure*}

Notice that the Laue condition\cite{kittel04, Ashcroft} implies that the constructive interferences will occur provided that the change in wave vector between the incident and reflected waves was a vector of the reciprocal lattice. Therefore the Bragg scattering condition is $|\vec{k}|=\frac{|\vec{G}|}{2\sin{\theta}}$. In the case 1D ($\theta=\pi/2$), the wavevectors are collinear and then, $|\vec{k}|=|\vec{G}|/2$.

\subsection{Design of the discrete materials with prescribed scattering properties}

\label{sec:op}

The optimization procedure consists of prescribing the scattering characteristics by using the structure factor and constructing configurations of multiple scatterers that give rise to these targeted value. In contrast to the real-space methods, we target information in reciprocal space to construct configurations whose structure factor exactly matches the candidate structure factor for a set of wavelengths. 

The objective function is based on the structure factor $S(\vec{G})$. For a given limit of wave vectors, fixing the optimization domain $\Omega$ in the reciprocal space, the structure factor should have a target value, $S_{op}$, for all the wave vectors in $\Omega$. Then, the objective functions will be based on the following expression
\begin{eqnarray}
\phi(\vec{r}_1,\dots,\vec{r}_N)=\sum_{\vec{G}\in\Omega}\left(S(\vec{G})-S_{op}\right),
\label{eq:potential}
\end{eqnarray}
and a standard deviation function
\begin{eqnarray}
\sigma=\sqrt{\frac{1}{N-1}\sum_{\vec{G}\in\Omega}|S(\vec{G})-\mu|^2},
\end{eqnarray}
where $\mu=\frac{1}{N}\sum_{\vec{G}\in\Omega}S(\vec{G})$ is the mean value in the target range of wave vectors. The optimization algorithm will look for configurations of multiple scatterers $\vec{r}_i$ that simultaneously minimizes the potential and the standard deviation function.

\subsection{Different kind of materials based on $S(\vec{G})$. Examples in 1D}
The interest of this work is focused on stealth materials, which are built from a configuration of multiple scatterers whose structure factor is exactly zero for some set of wavelengths. In the low-$k$ limit, several disordered systems, known as hyperuniform\cite{Torquato02, Torquato03, Florescu09, Man13a, Man13b, Torquato15, Torquato16, Gkantzounis17} materials, nearly suppress all scattering. These systems have the property that $\lim_{k\rightarrow0}S(\vec{G})=0$, i.e., infinite-wavelength density fluctuations vanish. Following this definition, we can see that periodic , i.e. crystalline  configurations, are by definition, in the Born approximation, hyperuniform since they suppress scattering for all wavelengths except those associated with Bragg scattering. It is worth noting here that other kind of materials can be defined by using the structure factor. For example, equiluminous materials scatter waves with the same intensity for a set of wave vectors. The structure factor for this class of materials is simply a constant for a set of wave vectors. Subsets of equiluminous materials include super-ideal gases  $S(\vec{G})=1$ as well as the stealth materials $S(\vec{G})=~0$. Super-ideal gases are single configurations of multiple scatterers whose scattering exactly matches that of an ensemble of ideal gas configurations, or Poisson point distributions, for a set of wave vectors, being $S(\vec{G})=cte.$

Figure \ref{fig:fig2} shows the 1D structure factor for several configurations made of $N=30$ point scatterers in a unit cell of length $L$. The first analysis is done for the periodic case in which the system has a periodicity of $L/N$ (see Fig. \ref{fig:fig2}(a)). In this case, the structure factor can be obtained analytically. If the Bragg condition is fulfilled the structure factor takes the value $N$ for $k_B=N\pi/L$. For wavelengths out of the Bragg condition, the structure factor is zero. This means that a periodic distribution of scatterers suppress scattering for all wavelengths except those associated with Bragg scattering, corresponding to a structure factor with a form of Dirac comb as shown in Fig. \ref{fig:fig2}(a). As soon as the structure becomes fully random, the periodic behavior of the structure factor is broken and random scattering is produced (see Fig. \ref{fig:fig2}(b)). The structure factor for a hyperuniform distribution of scatterers is shown in Fig. \ref{fig:fig2}(c).  The range of frequencies in which scattering is suppressed translates with the cancellation of the structure factor. As it can be seen, the suppression of the scattering is produced around $k=0$. Moreover, the structure factor peaks at wave vectors corresponding to the characteristics length scales of the periodic distribution, among them the first Bragg peak meaning that the system presents some hints of periodicity. This result is in agreement with those obtained in recent works showing that hyperuniform materials present isotropic band gaps.\cite{Florescu09, Man13a, Man13b}. Figure \ref{fig:fig2}(d) shows the results for a stealth material optimized in the range $k_B/5\leq |k| \leq k_B/1.5$. Effectively, the structure factor for such configuration of multiple scatterers is exactly zero for the prescribed set of wavelengths. This means that this structure completely suppress scattering of incident radiation for this set of wave vectors, and thus, is transparent at these wavelengths.


\section{Stealth materials for acoustic transparency}\label{sec:system}

\subsection{Stealth distribution of point scatterers}
In this Section, we design a 1D stealth configuration made of $N=25$ scatterers embedded in a fluid medium of length $L=1$ m for making the system transparent in a broad range of frequencies around the corresponding Bragg frequency of the equivalent periodic distribution of scatterers with periodicity $L/N$. The range of wave numbers to be optimized is $0.92 k_B<k<1.06k_B$. In particular, if we consider air as the fluid medium of the waveguide (with $c=340$ m/s the speed of sound), the range of frequencies to be optimized corresponds to the domain $f=[3910, 4504]$ Hz being $f_B=cN/L=4250$ Hz the Bragg frequency.  

\begin{figure}
\includegraphics[width=85mm]{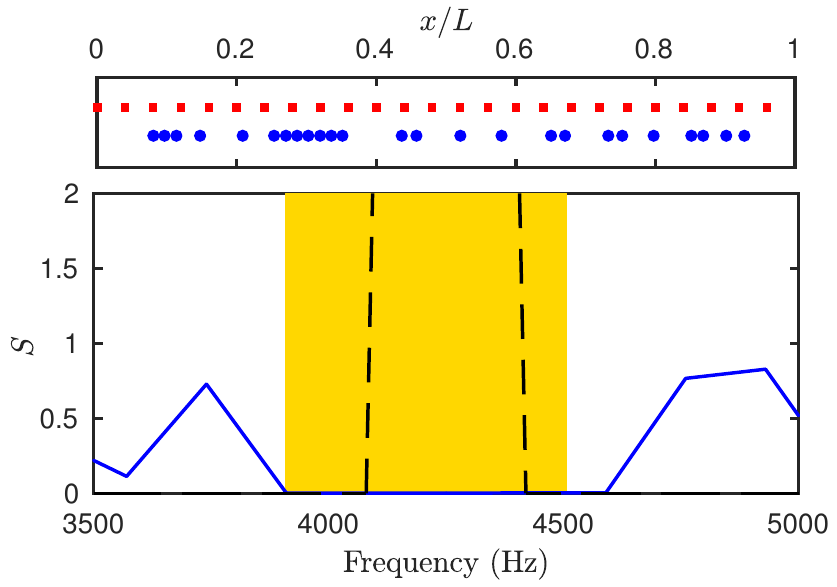}
\caption{(Color online) Structure factor for the periodic and stealth material. Upper panel represents both the periodic configuration [red squares ({\tiny \textcolor{red}{$\blacksquare$}})] and the stealth configuration [blue dots (\textcolor{blue}{$\bullet$})]. The positions of the scatterers in the stealth configurations are:  $x_n/L=[$0.083, 0.099, 0.116, 0.15, 0.211, 0.256, 0.273, 0.289, 0.305, 0.322, 0.338, 0.354, 0.439, 0.46, 0.523, 0.582, 0.653, 0.673, 0.735, 0.755, 0.8,  0.854, 0.871, 0.904, 0.93$]$. Lower panel shows the structure factor for the periodic distribution [black dashed line, (\textcolor{black}{$--$})] and for the stealth material [blue continuous line (\textcolor{blue}{$-$})]. Yellow area shows the frequency range of the optimization.}
\label{fig:stealth}
\end{figure}

The resulting configuration of multiple scatterers obtained from our optimization procedure is shown in the upper panel of Fig.~\ref{fig:stealth} with blue dots. The structure factor of the periodic case presents a peak at the Bragg frequency of the system, as shown in Fig. \ref{fig:stealth}. The optimized stealth structure has a minimum of the structure factor in the optimization range of  frequencies (yellow region). 

\subsection{Physical system: 1D waveguide with diaphragms}

\begin{figure}
\includegraphics[width=85mm]{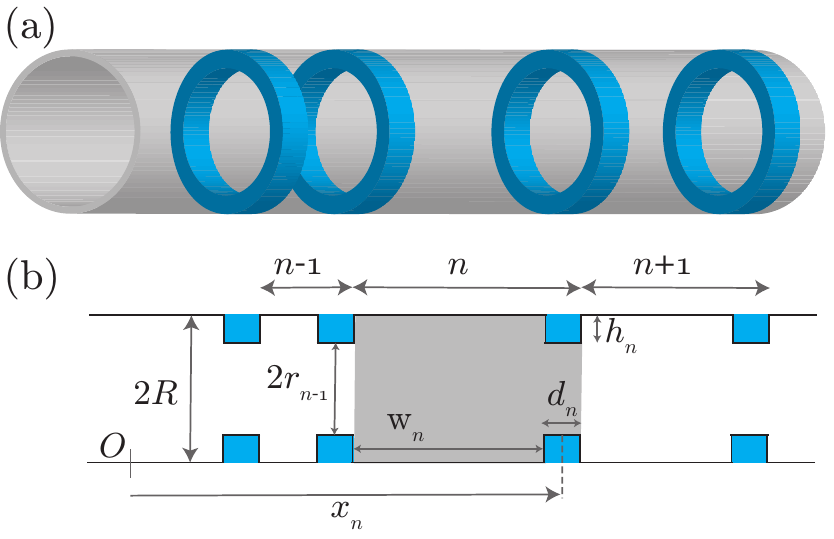}\\
\includegraphics[width=85mm]{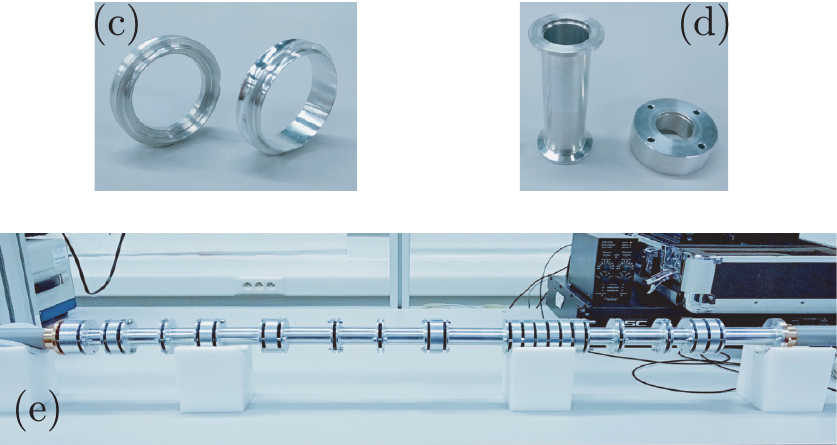}
\caption{(Color online) Schematics of both a three dimensional view (a) and the cross section (b) of the cylindrical waveguide along the $n$-th diaphragm. (c-e) Experimental set-up. (c) Connections: left diaphragme and flat connection for empty tube. (d) Separators. (e) Picture of the stealth material.}
\label{fig:fig3}
\end{figure}

To implement the stealth material we consider the propagation of sound waves within an air-filled cylindrical tube (with density and sound velocity of air defined as $\rho$ and $c$ respectively) of section $S=\pi R^2$ with diaphragms of length $d_n$ and section $s_n=\pi r_n^2$ placed along the waveguide at positions $\vec{x}_n$, as shown in Fig. \ref{fig:fig3}. The tube and the diaphragms are made of aluminium that can be considered acoustically rigid for the purpose of this work. The target frequency range is well below the first cutoff frequency of higher propagation modes in the waveguide, therefore the problem can be considered as one-dimensional. Moreover the size of the diaphragms can be considered small enough to be in the Born approximation.

\subsubsection{Theoretical model: Transfer Matrix Method}

The sound waves that propagate in the proposed acoustic structure are subject to the viscothermal effects on the walls which are taken into account by considering a complex expression for both the impedance and the wave number. In our case, we used the model of losses from Ref.  [\onlinecite{Theocharis14}], namely we replace the wave number and the characteristic impedance by the following expressions
\begin{eqnarray}
k_q     &=& \frac{\omega}{c}\left(1+\frac{(1+\imath)}{\sqrt{2}R_q/\delta}\frac{(1+(\gamma-1))}{\sqrt{P_r}}\right),\\
Z_q     &=& \frac{\rho c}{S_q}\left(1+\frac{(1+\imath)}{\sqrt{2}R_q/\delta}\frac{(1-(\gamma-1))}{\sqrt{P_r}}\right),
\end{eqnarray} 
where $\delta= \sqrt{2\mu/\rho\omega}$ is the viscous boundary layer thickness, $\mu=1.839\;10^{-5}$ kg m$^{-1}$ s$^{1}$ being the viscosity of air, $P_r=0.71$ the Prandtl number and $\gamma=1.4$ the heat capacity ratio of air. Notice that these expressions correspond for the wavevector and impedance of a cylindrical tube of radius $R_q$ and section $S_q$.

The transfer matrix between the two faces of the system, $\mathbf{T}$, extending from $x=0$ to thickness $x=L$, presenting $N$ diaphragms placed at position $\vec{x}_n$, relates the sound pressure, $p$, and normal acoustic flux, $v$, between its two faces by using
\begin{eqnarray}
\left[\begin{tabular}{c}
$p$\\
$v$
\end{tabular}\right]_{x=0}=
\mathbf{T}
\left[\begin{tabular}{c}
$p$\\
$v$
\end{tabular}\right]_{x=L},
\end{eqnarray}
where
\begin{eqnarray}
\mathbf{T}&=&
\left[\begin{tabular}{cc}
$T_{11}$ & $T_{12}$\\
$T_{21}$ & $T_{22}$
\end{tabular}\right]
=
\prod_{i=1}^N \mathbf{T_{w_i}} \mathbf{T_{d_i}},
\label{eq:totalmatrix}
\end{eqnarray}
with
\begin{eqnarray}
\mathbf{T_{w_i}}&=&\left[\begin{tabular}{cc}
$\cos({k_\mathrm{w}\mathrm{w}_i})$ & $-\imath Z_\mathrm{w}\sin({k_\mathrm{w}\mathrm{w}_i})$\\
$ -\frac{\imath}{Z_\mathrm{w}}\sin({k_\mathrm{w}\mathrm{w}_i})$ & $\cos({k_\mathrm{w}\mathrm{w}_i})$
\end{tabular}\right]
\end{eqnarray} 
and
\begin{eqnarray}
\mathbf{T_{d_i}}&=&\left[\begin{tabular}{cc}
$\cos({k_dd_i})$ & $-\imath Z_d\sin({k_dd_i})$\\
$ -\frac{\imath}{Z_d}\sin({k_dd_i})$ & $\cos({k_dd_i})$
\end{tabular}\right]
\end{eqnarray} 
describing the propagation through a waveguide of length $\mathrm{w}_i$ and a diaphragm of thickness $d_i$.

The reflection and transmission coefficients can be directly calculated from the elements of the matrix given in Eq. (\ref{eq:totalmatrix}) as \cite{Song00}
\begin{eqnarray}
\label{eq:T}
T&=&\frac{2e^{\imath k L}}{T_{11}+T_{12}/Z_\mathrm{w}+Z_\mathrm{w}T_{21}+T_{22}},\\
\label{eq:Rm}
R^-&=&\frac{T_{11}+T_{12}/Z_\mathrm{w}-Z_\mathrm{w}T_{21}-T_{22}}{T_{11}+T_{12}/Z_\mathrm{w}+Z_\mathrm{w}T_{21}+T_{22}},\\
\label{eq:Rp}
R^+&=&\frac{-T_{11}+T_{12}/Z_\mathrm{w}-Z_\mathrm{w}T_{21}+T_{22}}{T_{11}+T_{12}/Z_\mathrm{w}+Z_\mathrm{w}T_{21}+T_{22}},
\end{eqnarray}
where the superscripts $(+,-)$ denote the incidence direction, i.e., the positive and negative $x$-axis incidence respectively. We notice that, if the system was symmetric, then, $T_{11}=T_{22}$, and as a consequence, $R^+=R^-$. The reciprocal behaviour of the system can be seen from the fact that its determinant is one ($T_{11}T_{22}-T_{12}T_{21}=1$).

\subsubsection{Experimental set-up}
 
The experimental set-up was made of a cylindrical waveguide with radius $R=1.25$ cm, which cut-off frequency is around 8000 Hz well above the analyzed range of frequencies. Two types of pieces were built: the connections [Fig.~\ref{fig:fig3}(c)] and the separators [Fig.~\ref{fig:fig3}(d)]. Two types of connections were built, the ones with no diaphragm, in order to measure the properties of an empty waveguide, and the ones with a diaphragm, in order to obtain the scattering properties of the stealth material. The dimensions of the diaphragms were $d_n=d=2$ mm and $r_n=1.04$ cm, dimensions for which the Born approximation is fulfilled. The separators had different lengths in order to fit all the positions $x_n$ in the stealth material. A picture of the stealth material once the different pieces are properly assembled is shown in Fig.~\ref{fig:fig3}(e).

A loudspeaker is used to generate a plane wave field, and a single microphone was used to measure the transfer functions between the signal provided to the loudspeaker and the sound pressure at four locations in order to experimentally obtain the reflection and transmission coefficients of the stealth material (the $1$ microphone technique was used, see Ref.~[\onlinecite{Song00}]). On the other side of the tube, an anechoic termination with less than 5\% of reflection in the analyzed frequency range is used.

The stealth material is experimentally analyzed in the forward and in the backward incidence directions. To do that, we have to reverse the source and the anechoic termination in order to experimentally obtain all the scattering parameters, i.e., $R^+, R^-$ and $T$.

\section{Results}\label{sec:results}

In this Section, we analyze and present the theoretical results obtained from the TMM and the experimental results for the reflection, transmission and absorption coefficients of the designed stealth material. In addition to these results, we have used a numerical approach based on the Finite Element Method (FEM) to perform full wave numerical simulations in order to check that the possible evanescent coupling between scatterers is weak and can be neglected. In the numerical model the viscothermal losses were accounted for using the effective parameters defined above for each domain. An incident plane wave is used as excitation of the system and  perfectly matched layers were placed at the boundaries of the numerical domain to avoid unexpected reflections. The mesh is designed such as to ensure a maximum element size of $\lambda_B/20$ (being $\lambda_B=k_B^{-1}= L/N/\pi$).

\begin{figure}[h!]
\includegraphics[width=85mm]{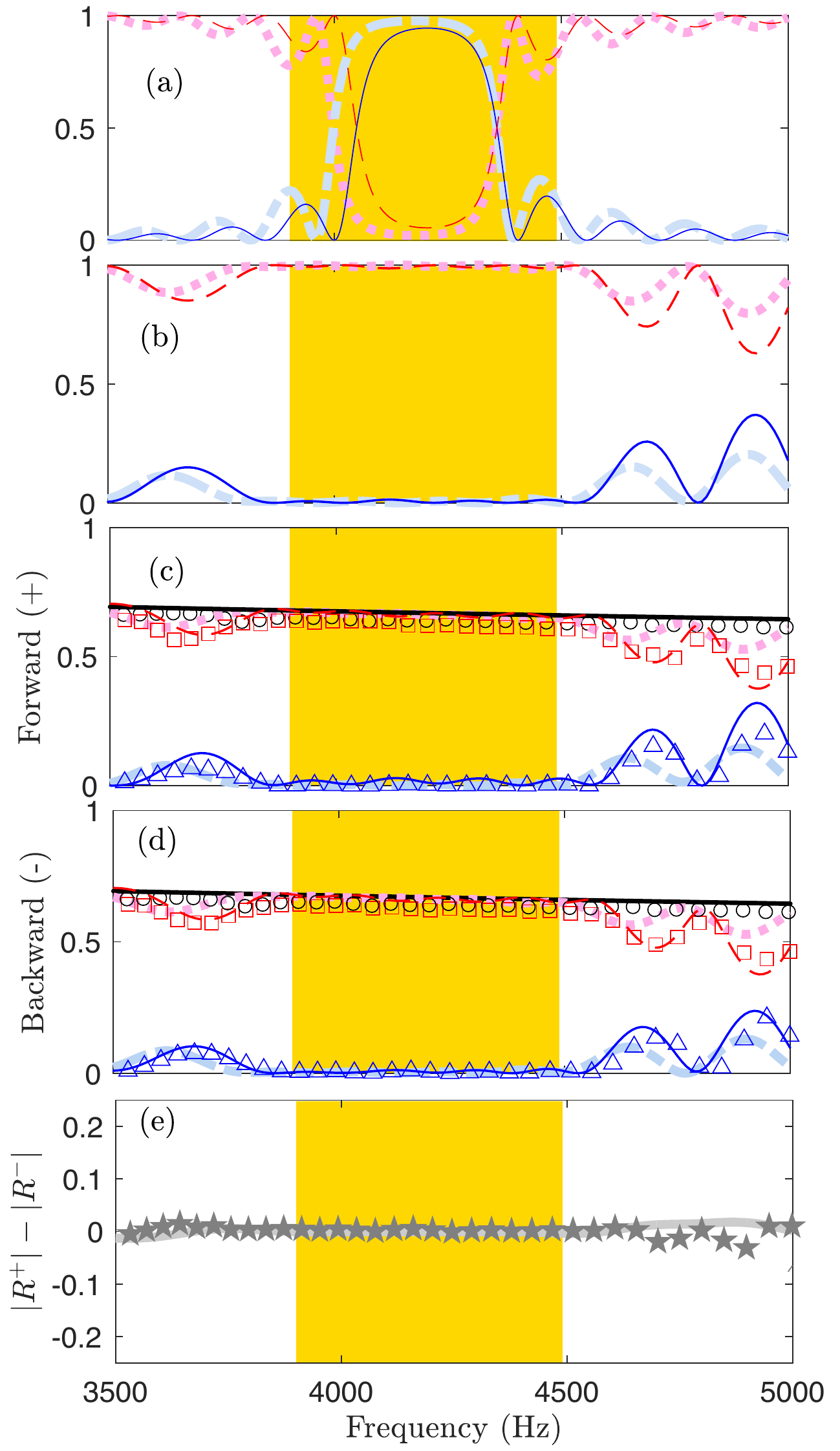}
\caption{(Color online) Analysis of the scattering produced by the stealth material. In the Figure the transmission, $T^2$, [reflection, $R^2$,] analytically and numerically calculated are represented by the shaded pink (\textcolor{redish}{$\rule{1mm}{2pt}\;\rule{1mm}{2pt}\;\rule{1mm}{2pt}$}) and red dashed (\textcolor{red}{$--$}) lines respectively [shaded blue (\textcolor{bluish}{$\rule{2mm}{2pt}\;\rule{0.5mm}{2pt}\;\rule{2mm}{2pt}$}) and blue  continuous (\textcolor{blue}{$-$}) lines respectively] and the experimental values are represented by red squares (\textcolor{red}{$\Box$}) [blue triangles (\textcolor{blue}{$\triangle$})]. (a) Scattering coefficients of (a) the periodic distribution in the lossless case, (b) the stealth material in the lossless case, (c) and (d) the stealth material in the lossy case excited from the forward and backward incidence directions respectively. Black (\textcolor{black}{$-$}) continuous line and black circles (\textcolor{black}{$\bigcirc$}) represent the transmission coefficient of the empty waveguide,  theoretical and experimental values respectively. (e) $|R^+|-|R^-|$ from the TMM (\textcolor{gray}{$\rule{2mm}{2pt}$}) and experimental results (\textcolor{gray}{$\bigstar$}).}
\label{fig:results}
\end{figure}

We start by the analysis of the lossless case. In Fig. \ref{fig:results}(a) we represent the scattering coefficients of the periodic distribution of scatterers. The Bragg scattering is shown in the target range of frequencies with an strong reduction of the transmission and an enhancement of the reflection, representing the band gap of the structure. The fact that we are dealing with finite structures explains the lobes of the reflection and transmission coefficients out of the band gap that corresponds to Fabry-Perot resonances.

Figure \ref{fig:results}(b) shows the scattering coefficients of the designed stealth material in the lossless case. The reflection coefficient is dramatically reduced and the transmission is almost one for the target range of frequencies. Therefore, the stealth material can be considered as transparent, even if $N$ scatterers are placed. For the frequencies out of the optimization range the system is not transparent. We notice that in this lossless case, $R^+$=$R^-$. The differences between the numerical and the analytical results out of the target frequencies shows that the evanescent coupling between the scatterers are not negligible. However in the optimization range, where the scattering is suppressed, the coupling is weak and the transparency of the system is shown by the numerics and analytics. Therefore, the optimization procedure is robust with respect to the evanescent coupling.

We analyze now the lossy case in order to show their effects on the scattering properties of the stealth material. Numerical, theoretical and experimental results are compared in Figs. \ref{fig:results}(c) and \ref{fig:results}(d). In this case, as the stealth material is not symmetric, $R^+\neq R^-$, but being reciprocal $T^+=T^-=T$. We start by analyzing the results for the system excited from the right side (forward direction, Fig. \ref{fig:results}(c)). In the optimized frequency band, the values of the reflection coefficient are nearly zero while the transmission coefficient has been reduced with respect to the lossless case. However the coefficients keep the flat behavior also shown in the lossless case. In order to check the transparency of the stealth material, we compare its transmission coefficients with the corresponding empty waveguide. In the optimized range of frequencies the transmission of the stealth material matches the transmission of the empty tube in very good agreement between numerics, analytics and experiments. As in the lossless case, out of the optimization range, the system is not transparent.


One of the properties of the structure factor is that $S(-\vec{G})=S(\vec{G})$, i.e. the scattered intensity should be the same in opposite directions even if the configuration of multiple scatterers conforming the stealth material is not symmetric, i.e., the reflection coefficient from both sides of the structure should be equal. Figure \ref{fig:results}(d) analyzes the same coefficients as Fig. \ref{fig:results}(c) but evaluated for the backward incidence direction, having a similar behavior as the opposite direction. Figure \ref{fig:results}(e) shows that the difference between the amplitudes of the reflection coefficients evaluated from each side of the stealth material is less than 3$\%$ in the whole frequency range. The slightly differences are due to the fact that the analyzed system is finite and the optimization procedure is based on an infinite one. Out of the target frequency range, the differences are bigger because the scattering is stronger than in the optimization range due to finite size effects of the system. However, the differences between the two sizes remain small than 3$\%$.

Finally, we want to point out that the reflection coefficient of all the analyzed systems in Figs. \ref{fig:results}(b-d) follows the same trends as the theoretical value of the structure factor, see Fig. \ref{fig:fig2}. This proves that the size of the diaphragms allows the system to work in the Born approximation.

\section{Conclusions}

By capturing a configuration of multiple scatterers in the direct space and Fourier transforming it to the reciprocal space, the structure factor has been used to design configurations of multiple scatterers with prescribed scattering characteristics in the Born approximation. The methodology proposed in this work allows us to engineer different kinds of materials: hyperuniform, steath, equiluminous and super-ideal materials. We have paid attention to the case of stealth materials, made from a configuration of multiple scatterers which structure factor is exactly zero for a target set of wavelengths. There is a main property of such materials: the broadband transparency in the target range of frequencies. The physical system used to check the theoretical predictions consists of an air-filled 1D waveguide with diaphragms. The positions of the centers of the diaphragms, which are small enough to fulfill the Born approximation, correspond to a stealth point distribution to avoid Bragg scattering and to achieve broadband transparency around this frequency. The system is analyzed analytically, numerically and experimentally along the two incidence directions showing very good agreement and reproducing the prescribed scattering properties of the system. Moreover and interestingly enough, we have proven that the optimization procedure is robust enough to be independent of the presence of losses. The optimization is also very robust to the finite size effects, showing that the property $S(\vec{G})=S(-\vec{G})$ is fulfilled with differences less than 3$\%$ between the reflection coefficients evaluated from each side of the structure in the whole range of frequencies. These preliminary results could be the basis for future applications in acoustics and more generally in wave physics. It is worth noting here that the methodology is general and can be applied for any type of wave. Extension to higher dimensions could provide very promising applications for the broadband control of waves.

\begin{acknowledgments}	
This work has been funded by RFI Le Mans Acoustique (R\'egion Pays de la Loire) in the framework of the APAMAS project, by the project HYPERMETA funded under the program \'Etoiles Montantes of the R\'egion Pays de la Loire as well as by the Ministerio de Econom\'ia y Competitividad (Spain) and European Union FEDER through project FIS2015-65998-C2-2-P. V. Romero-Garc\'{\i}a and L. M. Garc\'{\i}a-Raffi acknowledge the short term scientific mission (STSM) funded by the COST (European Cooperation in Science and Technology) Action DENORMS - CA15125.

\end{acknowledgments} 

%

\end{document}